\documentclass[journal,onecolumn,a4]{IEEEtran}
\usepackage{amsmath, amssymb, amscd, amsthm, amscd, amsfonts, parskip, xcolor, hyperref}
\usepackage{microtype}
\usepackage{graphicx}
\usepackage{multirow}
\usepackage{makeidx}
\usepackage{multirow}
\usepackage{multicol}
\newtheorem{de}{Definition}

\newtheorem{lem}{Lemma}
\newtheorem{thm}{Theorem}
\newtheorem{re}{Remark}

\begin{document}
\title{URDP: General Framework for Direct CCA2 Security from any Lattice-Based PKE Scheme}
\author{Roohallah Rastaghi}
\maketitle
\begin{center}
{r.rastaghi59@gmail.com}
\end{center}
\begin{center}
February 22, 2013\\
\end{center}

\begin{abstract}
Design efficient lattice-based cryptosystem secure against adaptive chosen ciphertext attack ({\sf IND-CCA2}) is a challenge problem. To the date, full CCA2-security
of all proposed lattice-based cryptosystems achieved by using a generic transformations such as either strongly unforgeable one-time signature schemes ({\sf SU-OT-SS}), or a
message authentication code ({\sf MAC}) and weak form of commitment. The drawback of these schemes is that encryption requires {\it separate encryption}.
Therefore, the resulting encryption scheme is not sufficiently efficient to be used in practice and it is inappropriate for many applications such as small ubiquitous computing devices with limited resources such as smart cards, ‎active RFID tags, wireless sensor networks and other embedded devices.

In this work, for the first time, we introduce an efficient universal random data padding ({\sf URDP}) scheme, and show how it can be used to construct a {\it direct} CCA2-secure encryption scheme from {\it any} worst-case hardness problems in (ideal) lattice in the standard model, resolving a problem that has remained
open till date. This novel approach is a {\it black-box} construction and  leads to the elimination of separate encryption, as it avoids using general transformation from CPA-secure scheme to a CCA2-secure one. IND-CCA2 security of this scheme can be tightly reduced in the standard model to the assumption that the underlying primitive is an one-way trapdoor function.
\end{abstract}
\section{Introduction} \label{sec1}
\quad Devising quantum computer will enable us to break public-key cryptosystems based on integer factoring ({\sf IF}) and discrete logarithm ({\sf DL}) problems\cite{17} \label{17}.
Under this future threat, it is important to search for secure PKEs based on the other problem. Lattice-based PKE schemes hold a great promise for post-quantum cryptography, as they
enjoy very strong security proofs based on worst-case hardness, relatively efficient implementations, as well as great simplicity and, lately, their promising potential as a platform
for constructing advanced functionalities.

The ultimate goal of public-key encryption is the production of a simple and efficient encryption scheme that is provably secure in a strong security model under a weak and reasonable
computational assumption. The accepted notion for the security of a public-key encryption scheme is semantically secure against adaptive chose ciphertext attack (i.e. IND-CCA2) \cite{13}
\label{13}. In this scenario, the adversary has seen the \textit{challenge ciphertext }before having access to the decryption oracle. The adversary is not allowed to ask the decryption
of the challenge ciphertext, but can obtain the decryption of {\it any} relevant cryptogram (even modified ones based on the challenge ciphertext). A cryptosystem is said to be CCA2-secure if the cryptanalyst fails to obtain any partial information about the plaintext relevant to the challenge ciphertext.
\subsection{Related work} \label{subsec 1.1}
\quad In order to design CCA2-secure lattice-based encryption schemes, a lot of successes were reached. There are two approach for constructing CCA2-secure lattice-based cryptosystems in the standard model. Existing CCA2-secure schemes exhibit various incomparable tradeoffs between key size and error rate.

{\bf-}{\it CCA-secure cryptosystem based on lossy trapdoor functions}.
Peikert and Waters \cite{11} \label{11} showed for the first time how to construct CCA2-secure encryption scheme from a
primitive called a {\it lossy} {\sf ABO} trapdoor function family, along with a {\sf SU-OT-SS}. They showed how to construct this primitive based on the learning with error
({\sf LWE}) problem. This result is particularly important as it gives for the first time a CCA-secure cryptosystem based on the worst-case hardness of lattice problems. It has
public-keys of size ${\cal O} (n^2)$ bits and relies on a quite small {\sf LWE} error rate of $\alpha = {\cal O}(1/n^4)$. Subsequently, Peikert \cite{12} \label{12}showed
how to construct a correlation-secure trapdoor function family from the {\sf LWE} problem, and used it within the Rosen-Segev scheme \cite{15} \label{15}to obtain another lattice-based CCA-secure scheme. Unfortunately, the latter
scheme also suffers from long public-key and ciphertext length of ${\cal O} (n^3 )$ bits, but uses a better error rate of ${\cal O}(1/n)$ in the security parameter $n$, even if applied in the Ring-{\sf LWE} setting.
Recently, Micciancio and Peikert \cite{10} \label{10}give new methods for generating simpler, tighter, faster and smaller trapdoors in cryptographic lattices to achieve a CCA-secure cryptosystem. Their construction give a CCA-secure cryptosystem
that enjoys the best of all prior constructions, which has ${\cal O} (n^2)$ bit public-keys, uses error rate ${\cal O}(1/n)$.
Recently, Steinfeld et al. \cite{18} \label{18}introduced the first CCA2-secure variant of the NTRU \cite{9}\label{9} in the standard model with a provable security from worst-case problems in \textit{ideal lattices}.
They construct a CCA-secure scheme using the lossy trapdoor function, which they generalize it to the case of $(k - 1)$-of-$k$-correlated input distributions.

{\bf-}{\it CCA-secure cryptosystem based on IBE}. More constructions of IND-CCA2 secure lattice-based encryption schemes can be obtained by using the lattice-based
selective-ID secure identity-based encryption ({\sf IBE}) schemes of \cite{1,2,3,4,7,14,16,19} \label{1}\label{2}\label{3}\label{7}\label{14}\label{16}\label{19}\label{4}within the generic constructions of \cite{5,6}\label{5}\label{6}, and a {\sf SU-OT-SS} or commitment scheme.

\quad All the above schemes use generic transformations from CPA to CCA2 security in the standard model, e.g., Dolev et al. approach \cite{8}, Canetti et al. paradigm \cite{6} or Boneh et al. approach \cite{5}. They typically involve either a {\sf SU-OT-SS} or a {\sf MAC} and commitment schemes to make the ciphertext authentic and non-malleable. So, the resulting encryption scheme requires {\it separate encryption} and thus, it is not sufficiently efficient to be used in practice and inappropriate for many applications such as small ubiquitous computing devices with limited resources such as smart cards, ‎active RFID tags, wireless sensor networks and other embedded devices.

\quad Till date, there is no generic {\it direct} transformation from {\it any} lattice-based one-way trapdoor cryptosystem (i.e., worst-case hardness problem in lattice) to a CCA2-secure one. In this work, for the first time, we show how to construct a CCA2-secure cryptosystem directly based on the worst-case hardness problems in lattice, resolving a problem that has remained open till date.
\subsection{Our contributions}
Our approach has several main benefits:
  \begin{itemize}
  \item It introduce a new generic asymmetric padding-based scheme. The main novelty is that our approach can be applied to {\it any} conjectured (post-quantum) one-way trapdoor cryptosystem.
  \item Our approach yields the first known {\it direct} CCA2-secure PKE scheme from worst-case hardness problems in lattice.
    \item The proposed approach is a "black-box" construction, which making it more efficient and technically simpler than those previously proposed. The publick/secret keys are as in the original scheme and the encryption/decryption complexity are comparable to the original scheme.
    \item This novel approach leads to the elimination of using generic transformations from CPA-secure schemes to a CCA2-secure one.
   \item Our CCA2-security proof is tightly based on the assumption that the underlying primitive is a trapdoor one-way function. So, the scheme's {\it consistency} check can be directly implemented by the {\it simulator} without having access to some external gap-oracle as in previous schemes \cite{1,2,3,7,10,11,12,14,16,18,19}.\label{1-1}\label{2-1}\label{3-1}\label{7-1}\label{10-1}\label{11-1}\label{12-1}\label{14-1}
       \label{16-1}\label{18-1}\label{19-1}Thus, our proof technique is fundamentally different from all known approaches to obtain CCA2-security in the lattice-based cryptosystems.
  \item Additionally, this scheme can be used for encryption of {\it arbitrary-length} long messages without employing the hybrid encryption method and symmetric encryption.
  \end{itemize}
{\bf Organization.} The rest of this manuscript is organized as follows: In the following section, we briefly explain some notations and definitions. Then, in Section
3, we introduce our proposed scheme. Security and performance analysis of the proposed scheme will be discussed in Section 4.
\section{Preliminary}\label{sec2}

\subsection{Notation}
We will use standard notation. If $x$ is a string, then $\left| x \right|$ denotes its length. If $k \in \mathbb{N}$, then $\left\{ {0,\,1} \right\}^k$ denote the set of \textit{k}-bit strings,
$1^k$ denote a string of $k$ ones and $\left\{ {0,\,1} \right\}^*$ denote the set of bit strings of finite length. $y \leftarrow x$ denotes the assignment to \textit{y }of the value
\textit{x}. For a set $S$, $s \leftarrow S$ denote the assignment to $s$ of a uniformly random element of $S$. For a deterministic algorithm ${\cal A}$, we write
$x\leftarrow {\cal A}^{\cal O} (y,\,z)$ to mean that \textit{x} is assigned the output of running ${\cal A}$ on inputs \textit{y} and \textit{z}, with access to oracle ${\cal O}$.
We denote by ${\rm Pr}[E]$ the probability that the event $E$ occurs. If $a$ and $b$ are two strings of bits, we denote by $a\|b$  their concatenation. The bit-length of $a$ denoted
by ${\sf Len}(a)$, ${\sf Lsb}_{x_1}(a)$ means the right $x_1$ bits of $a$ and ${\sf Msb}_{x_2}(a)$ means the left $x_2$ bits of \textit{a}.

\subsection{Definitions}
\begin{de} [{\bf Public-key encryption scheme}] A public-key encryption scheme $($PKE$)$ is a triple of probabilistic polynomial time
 $($PPT$)$ algorithms $(${\sf Gen},\,{\sf Enc},\,{\sf Dec}$)$ such that:
\begin{itemize}
\item {\sf Gen} is a probabilistic polynomial-time key generation algorithm which takes a security parameter $1^n$ as input and
outputs a public key $pk$ and a secret-key $sk$. We write $(pk,sk) \leftarrow {\sf Gen}(1^n )$. The public key specifies the
message space ${\cal M}$ and the ciphertext space ${\cal C}$.
\item {\sf Enc} is a $($possibly$)$ probabilistic polynomial-time encryption algorithm which takes as input a public key \textit{pk,} a
$m\in {\cal M}$ and random coins $r$, and outputs a ciphertext $C \in {\cal C}$. We write ${\sf Enc}(pk,m; r)$ to
indicate explicitly that the random coins \textit{r} is used and ${\sf Enc}(pk,m)$ if fresh random coins are used.
\item {\sf Dec} is a deterministic polynomial-time decryption algorithm which takes as input a secret-key \textit{sk} and a ciphertext
$C \in {\cal C}$, and outputs either a message $m \in {\cal M}$ or an error symbol $\bot$. We write $m \leftarrow{\sf Dec}(C,\,sk)$.
\item $($Completeness$)$ For any pair of public and secret-keys generated by Gen and any message $m \in {\cal M}$ it holds that
${\sf Dec}(sk,\,{\sf Enc}(pk,m;r))=m$ with overwhelming probability over the randomness used by {\sf Gen} and the random coins \textit{r} used by {\sf Enc}.
\end{itemize}
\end{de}
\begin{de}[{\bf Padding scheme}] Let $\nu, \rho, k$ be three integers such that ${\nu}+{\rho} \leq k$. A padding scheme ${\rm \Pi}$ consists of two mappings $\pi : \{0, 1\}^{\nu} \times \{0, 1\}^{\rho} \rightarrow \{0, 1\}^{k}$ and $\hat{\pi} : \{0,1\}^{k} \rightarrow \{0,1\}^{\nu} \times \{0,1\}^{\rho} \cup \{\bot\}$ such that $\pi$ is injective and the following consistency requirement is fulfilled:
\[
\forall m \in \{0, 1\}^{\nu}, r \in \{0, 1\}^{\rho} : \hat{\pi}(\pi(m,r))=m .
\]
\end{de}
\begin{de} [{\bf CCA2-security}] A public-key encryption scheme PKE is secure against adaptive chosen-ciphertext attacks $($i.e. IND-CCA2$)$ if the advantage of any two-stage PPT
adversary ${\cal A} = ({\cal A}_1 ,\,{\cal A}_2 )$ in the following experiment is negligible in the security parameter $k$:

\quad \quad \quad ${\bf Exp}_{{\rm PKE}, {\cal A}}^{cca2}(k)$:

\quad \quad \quad\quad $(pk, sk)\leftarrow {\sf Gen}(1^k)$

\quad \quad \quad\quad $(m_0, m_1, {\sf state})\leftarrow {\cal A}_{1}^{{\sf Dec}(sk, .)} (pk) \quad {\rm s.t.} \quad |m_0|=|m_1|$

\quad \quad \quad \quad $b\leftarrow \{0, 1\}$

\quad \quad \quad \quad $C^{*}\leftarrow {\sf Enc}(pk, m_b)$

\quad \quad \quad \quad $b'\leftarrow {\cal A}_{2}^{{\sf Dec}(sk, .)}(C^{*}, {\sf state})$

\quad \quad \quad \quad  if $b= b^{'}$  return 1, else return 0.

The attacker may query a decryption oracle with a ciphertext $C$ at any point during its execution, with the exception that ${\cal A}_2 $ is not allowed to query
${\rm Dec} (sk,\,.)$ with $ C^{*} $. The decryption oracle returns $ b^{'}  \leftarrow {\cal A}_2^{{\rm Dec} (sk,\,.)} (C^{*} ,{\sf state})$.
The attacker wins the game if $b = b'$ and the probability of this event is defined as $ \Pr [{\rm Exp} \,_{{\rm PKE} ,\,{\cal A}}^{cca2} \,(k)] $.
 We define the advantage of $ {\cal A}$ in the experiment as

\[
{\sf Adv}_{{\rm PKE} ,\,{\cal A}}^{\rm IND-CCA2} \,(k) = \left| \Pr [{\rm Exp} \,_{{\rm PKE} ,\,{\cal A}}^{cca2} \,(k) = 1] - \frac{1}{2} \right|.
\]
\end{de}
\section{The proposed cryptosystem}
\quad In this section, we introduced our proposed CCA2-secure encryption scheme. Our scheme is a precoding-based algorithm which can transform any one-way trapdoor cryptosystem to a CCA2-secure one in the standard model. Precoding includes a permutation and pad some random obscure-data to the message bits.

\subsection{The proposed idea}
\quad Let we can decide to encrypt message $m \in \{0,\,1\}^n$. At first, we perform a random encoding to the message bits. To do this, we uniformly choose $r=(r_1, \ldots, r_k) \in_R \{0,1\}^k$
with $k\ll n$ at random, and, suppose ${\sf wt}(r)=h$ be the its Hamming weigh. If $ n/h$ is an integer, then we can divide $m$ into $h$ blocks. Otherwise, in order to divide $m$ into
$h$ blocks, we must pad a random binary string (${\sf {\sf RBS}}$) with length $h\,.\left\lceil n / h \right\rceil-n$ to the right of $m$. In each cases, we can divide $m$ into $h$ blocks $d_1 \| d_2 \| \ldots \| d_h$ with equal binary length $v = \left\lceil n/h \right\rceil$ where $d_h = {\sf Lsb}_{(n-(h-1)\,.\left\lceil n/h \right\rceil )}\,(m) \|{\sf RBS}$.
Therefore, if $h\mid n$, then ${\sf RBS}= \varphi$ (the empty set) and $d_h ={\sf Lsb}_{(n-(h-1)\,.\left\lceil n/h \right\rceil )}\,(m)$, else, ${\sf RBS}$ is a random block with binary length $ h\,.\left\lceil n/h \right\rceil-n$ and we have $d_h={\sf Lsb}_{(n-(h-1)\,.\left\lceil n/h \right\rceil)}\,(m) \|{\sf RBS}$.

\quad Now, we perform a random permutation and pad some random obscure blocks (${\sf ROB}$s) with equal binary length $s$ into the message blocks $d_i ,\,1 \le i \le h$ using padding scheme
$\pi:\,\,\,\left\{ {0,\,1} \right\}  \times \left\{ {0,\,1} \right\}^v  \rightarrow \left\{ {0,1} \right\}^v  \times \left\{ {0,1} \right\}^s$, which can be defined as follows:

\[
\pi(r_i ,\,d_i ) = d^{'}_i  = \left\{
  \begin{array}{lll}
    d_{\sum_{j=1}^{i} r_j} \,\,\,\, {\rm if}\,\,\,r_i  = 1 \\
\\
{\sf{ROB}}\quad\,\,\,\,\,\,\,{\rm if}\,\,\,r_i  = 0
  \end{array}
\right.,\,\, 1\leq i\leq k.
\]

Notice that in order to prevent excessive increase in the message length, we can choose $s$ small enough. The message $m^{'} = (d^{'}_1 \| d^{'}_2 \|\ldots\| d^{'}_k )$ is called
encoded message. We summarize encoding process in algorithm 1.

{\bf Algorithm 3.1: Random Encoding Algorithm.}\label{alg1}

{\it Input:} $m=(m_1, \ldots, m_n)$, $r\in_R \{0,1\}^k$ with $n\gg k$ .

{\it Output:} Encoded message $m' = (d^{'}_1 \| d^{'}_2 \|\ldots \| d^{'}_k )$.

{\it SETUP:}‎
\begin{enumerate}
\item $h \leftarrow {\sf wt}(r)$.
\item If ‎$h\mid n$ then ‎$v \leftarrow n/h$;

 else $v \leftarrow \left\lceil n/h\right\rceil$ and choose a {\sf RBS} with‎ binary length $h\cdot\left\lceil n/h\right\rceil-n $, and

\quad \quad $m \leftarrow (m_1, \ldots ,m_n\|\underbrace{\sf R\,\,\, B\,\,\, S}_{h\cdot\left\lceil n/h\right\rceil-n })$.
\item Divide $m$‎ ‎ into $h$ blocks ‎$(d_1 \| d_2 \|\,\ldots\| d_h )$ with equal ${\sf Len}(d_i)=v, \,1\leq i\leq h$.
\end{enumerate}
{\it PERMUTATION AND PADDING:‎}
\begin{enumerate}
\item Uniformly choose integer $s$ at random‎.  ‎
\item For $i=1$ to $k$‎ do‎;

\quad \quad if $r_i=1$ then $d^{'}_i\leftarrow d_{\sum_{j = 1}^{i} {r_j } }$,

\quad \quad else $d_i^{'}\leftarrow{\sf ROB}$ with binary length $s$.

Return‎ $m^{'} = (d^{'}_1 \| d^{'}_2 \|\,\ldots\| d^{'}_k )$.
\end{enumerate}

\quad We illustrate algorithm (\ref{alg1}) with small example. Suppose $m = (m_1, \ldots ,m_{1117} )$ and $r = (0,1,0,1,1,0,1,0,1,1,1,0,1,1,1,0,1,0)$.

\textit{SETUP:}

We have $\left|m\right| = n = 1117$, $k=18$ and $h = \sum_{i = 1}^k {r_i }  = 11$. Since $11\nmid 1117$ so we must pad a {\sf RBS} with binary length $h\,.\left\lceil n/h
\right\rceil-n = 5$ to the right of $m$. If we uniformly chose ${\textcolor[rgb]{0.00,0.00,1.00}{1,0,1,1,0}}$ at random, we have $ m = (m_1 , \ldots ,m_{1117} ,\,\underbrace {\textcolor[rgb]{0.00,0.00,1.00}{{1,0,1,1,0}}}_{h\,.\left\lceil n/h \right\rceil-n})$. Since $h = 11$, the algorithm divides $m$ into 11 blocks with equal length
$v = \left\lceil n/h \right\rceil=102$. \sloppy We have $m = (\underbrace {m_1 ,\ldots ,\,m_{102} }_{d_1 }\| \underbrace {m_{103} ,\ldots ,m_{204} }_{d_2 }\| \ldots\,\| \underbrace
{m_{1020} , \ldots ,m_{1117} ,{\textcolor[rgb]{0.00,0.00,1.00}{1,0,1,1,0}}}_{d_{11} })$, where ${\sf Lsb}_{(n-(n-1)\cdot \lceil n/h\rceil )} \,(m) = {\sf Lsb}\, _{97}\,(m) = m_{1020} , \ldots ,m_{1117}$.

\textit{PERMUTATION AND PADDING:}

Firstly, we choose random integer \textit{s}, say $s = 4$. We have

$r_1=0$, thus $d^{'}_1  \leftarrow {\sf ROB} \# 1 = {\textcolor[rgb]{1.00,0.00,0.00}{(0,1,1,0)}}$, where ${\textcolor[rgb]{1.00,0.00,0.00}{(0,1,1,0)}}$ is randomly chosen by
algorithm 3.1.

$r_2=1 $, thus $d^{'}_2  \leftarrow d_{\sum_{j = 1}^2 {r_j } }  = d_1$.

$r_3  = 0$, thus $d^{'}_3  \leftarrow {\sf ROB} \# 2 = {\textcolor[rgb]{1.00,0.00,0.00}{(1,0,1,0)}}$, where ${\textcolor[rgb]{1.00,0.00,0.00}{(1,0,1,0)}}$ is randomly chosen by
algorithm 3.1.\\
\vdots\\
$r_{17} = 1$, thus $d^{'}_{17}  \leftarrow d_{\sum_{j = 1}^{17} {r_j }} = d_{11}$.

$r_{18}  = 0$, thus $d^{'}_{18} \leftarrow {\sf ROB} \# (k-h)=7 = {\textcolor[rgb]{1.00,0.00,0.00}{(0,0,1,0)}}$, where ${\textcolor[rgb]{1.00,0.00,0.00}{(0,0,1,0)}}$ is randomly chosen by algorithm 3.1.

$ l-h $ {\sf ROB} blocks with equal length $s=4$ are combined with the message blocks $d_i ,\,1 \le i \le h$, to produce the encoded message $ m^{'}  = (d^{'}_1 \| d^{'}_2 \| \ldots \| d^{'}_k )$. In the final, the algorithm outputs $m^{'}$ as $m^{'} = (\underbrace {{\textcolor[rgb]{1.00,0.00,0.00}{0,1,1,0}}}_{d^{'}_1 }\| \underbrace {m_1 ,\ldots,\,m_{102}}_{d^{'}_2}\|
\underbrace {{\textcolor[rgb]{1.00,0.00,0.00}{1,0,1,0}}}_{d^{'}_3 }\| \ldots\| \underbrace {m_{1020} , \ldots ,m_{1117} ,{\textcolor[rgb]{0.00,0.00,1.00}{1,0,1,1,0}}}_{d^{'}_{17}}\,\|\,\|
\underbrace {{\textcolor[rgb]{1.00,0.00,0.00}{0,0,1,0}}}_{d^{'}_{18}}).$

\quad As we see, the \textit{length} and the \textit{position} of the message blocks $d_i$ are correlated to the \textit{number} and the \textit{position} of the random bits
$r_i=1$ respectively, and completely random.
\subsection{The proposed scheme}
\quad Now, we are ready to define our proposed encryption scheme. Given a secure lattice-based encryption scheme ${\rm \Pi}_{\rm lbe}=({\sf Gen}_{\rm lbe},{\sf Enc}_{\rm lbe}, {\sf Dec}_{\rm lbe})$, we construct a IND-CCA2 secure encryption scheme ${\rm \Pi_{cca2}}=({\sf Gen_{cca2}}, {\sf Enc_{cca2}}, {\sf Dec_{cca2}})$ as follows. This scheme can be used for encryption of {\it arbitrary-length} long messages.
{\bf System parameters.} $n, k \in \mathbb{N}$, where $n\gg k$.\\
{\bf Key generation.} Let ${\sf Gen}_{\rm lbe}$ be the Lattice-based key generator. On security parameter $1^k$, the generator ${\sf Gen_{cca2}}$ runs ${\sf Gen}_{\rm lbe}(1^k)$ to obtain
\[
sk= sk_{\rm lbe} \quad {\rm and} \quad pk= pk_{\rm lbe}.
\]

{\bf Encryption.}
To encrypt message $m \in \{0,\,1\}^n$ with $n\gg k$, ${\sf Enc_{cca2}}(pk, m)$ works as follows.
\begin{itemize}
\item Uniformly chooses $ r\in_R \{0,1\}^k$ at random and computes its Hamming weight ${\sf wt}(r)=h$.
\item Randomly chooses small integer $s$ and executes algorithm (\ref{alg1}) for generate encoded message $m^{'}  = (d^{'} _1 \| d^{'} _2 \| \ldots \| d^{'} _l)$ from message $m$.
\item Suppose $y$ be the corresponding decimal value of $ m^{'}$. Computes
\end{itemize}

\[
C_1  = y\cdot h , \quad C_2={\sf Enc}_{\rm lbe}(pk, r)
\]
and outputs the ciphertext $C=(C_1, C_2)$.

\quad To handle CCA2-security and non-malleability related issues, we strictly correlate the message bits $m_i,\, 1\leq i\leq n$ to the randomness $r$ via encoding process.
The value of $y$ also correlates to the random binary string $r$ via its Hamming weight $h={\sf wt}(r)$. So, the CCA2 adversary for extract the message blocks $d_i$
from $C_1$ must first recover {\it exactly} the same random binary string $r$ from lattice-based cryptosystem which is impossible, if the underlying lattice-based one-way trapdoor cryptosystem
be secure.

{\bf Decryption.}
${\sf Dec_{cca2}}(sk, C)$ for extract message $m$ performs the following steps.

\begin{itemize}
	\item Computes random binary vector $r$ as $r = {\sf Dec}_{\rm lbe}(C_2, sk)$ and $h = \sum_{i = 1}^k {r_i }$.
	\item Computes $y = {C_1}/h$.
	\item Checks whether
 \begin{equation}
 {\sf Len}(y)\stackrel{?}{=} h\cdot \left\lceil {n/h} \right\rceil
 \label{eq1}
 \end{equation}
  holds, and rejects if not ({\it consistency} check). If (\ref{eq1}) hold, computes $v = \left\lceil {n/h} \right\rceil$ and binary coded decimal ({\sf BCD}) $m'$ of $y$.
  \item Computes $s = (\left|m'\right|- hv)/(k-h)$ and rejects the ciphertext if $s$ is not an integers (verify whether the padding information is correct or not).
 \item The {\it lengths} and {\it position} of the message/{\sf ROB} blocks are explicit,
therefore, ${\sf Dec_{cca2}}$ simply can separate {\sf ROB} blocks from encoded message $m'$ and extract message blocks $d_i ,\,\,1 \le i \le h$ with the following algorithm.
\end{itemize}
{\bf Algorithm 3.2: Message Extractor.}\label{alg2}\\
{\it Input:} $r=(r_1, \ldots, r_k)$, integers $h, v, s$ and encoded message $m'$.\\
{\it Output:} Retrieved message $m=(d_1\|d_2\| \ldots \|d_h)$

\begin{enumerate}
\item For $i=1$ to $k$ do

\quad If $r_i=0$, then $m'\leftarrow {\sf Lsb}_{(\left|m'\right|-s)} (m')$,

\quad else $d_{\sum_{j=1}^{i} r_j}\, \leftarrow {\sf Msb}_v (m')$ and $m'\leftarrow {\sf Lsb}_{(\left|m'\right|-v)} (m')$;

\item $m \leftarrow (d_1\|d_2\| \ldots \|d_{\sum_{j=1}^{k} r_j})$, where $\sum_{i=1}^{k} r_i = h$.

\item If $h\nmid n$, then $m \leftarrow {\sf Msb}_n (m)$ (remove right $(h.\lceil n/h \rceil -n)$ bits of $m$).
\end{enumerate}
Return "$m$".
\section{Security and performance analysis}
\label{sec4}

\subsection{Security analysis}
\quad In this subsection, we proof the CCA2-security of the proposed cryptosystem which is built using the pre-coding approach with a secure lattice-based encryption scheme.
\begin{thm}: Let ${\rm \Pi}_{\rm lbe}=({\sf Gen}_{\rm lbe},{\sf Enc}_{\rm lbe}, {\sf Dec}_{\rm lbe})$ be a secure lattice-based encryption scheme, then the proposed scheme is CCA2-secure in the standard model.
\end{thm}
In the proof of security, we exploit the fact that for a well-formed ciphertext,‎ we can recover the message if we know the randomness $r$ that was used to create the ciphertext.‎

\textbf{Proof:} Suppose that $ C^*  = (C_1^*,C_2^* )$ be the challenge ciphertext. Let $S_i$ be the event that the adversary ${\cal A}$ wins in Game $i$. Here is the sequence of games.

{\bf Game 0.} We define Game 0 which is an interactive computation between an \textit{adversary} $ {\cal A}$ and a \textit{simulator}. This game is usual CCA2 game used to define
CCA2-security, in which the simulator provides the adversary's environment. \\
Initially, the simulator runs the key generation algorithm and gives the public-key to the adversary.
The adversary submits two messages $m_0, m_1$ with $|m_0|= |m_1|$ to the simulator. The simulator chooses $b \in \{ 0,1\}$ at random, and encrypts $m_b$, obtaining the challenge
ciphertext $C^*  = (C_1^*,C_2^* )$. The simulator gives $C^*$ to the adversary. We denote by $r^{*}$, $h^{*}={\sf wt}(r^{*})$,
$v^{*}=\left\lceil n/h^{*}\right\rceil$, $s^*$ and $y^{*}= {\sf DV}(m'^{*})$ where
\begin{equation}\label{eq2}
m'^{*}={\sf Encode}(m_b, r^*,s^*)
\end{equation}
the corresponding intermediate quantities computed by the encryption algorithm, where {\sf DV} means the decimal value.
The only restriction on the adversary's requests is that after it makes a challenge request, the subsequent decryption requests must not be
the same as the challenge ciphertext. At the end of the game, the adversary ${\cal A}$ outputs $ \tilde b \in \{ 0,\,1\}$. Let $S_0$ be the
event that $\tilde b=b$. Since Game 0 is identical to the CCA2 game we have that

\[
\left| \Pr [S_0]- \frac{1}{2} \right|={\sf Adv}_{{\rm \Pi},\, {\cal A}}^{cca2} \,(k).
\]

and, our goal is to prove that this quantity is negligible.

{\bf Game 1.} Define Game 1 as identical with Game 0, except that $h = h^{*}$.
\begin{lem} There exists an efficient adversary ${\cal A}_1$ such that:
\begin{equation}\label{eq3}
\left| \Pr [S_1]-\Pr [S_0] \right| \le {\sf Adv}_{{\rm \Pi}, \,{\cal A}_1}^{\rm lbe} (k).
\end{equation}
By the assumption that the lattice-based encryption scheme is secure, we have that ${\sf Adv}_{{\rm \Pi}, \,{\cal A}_1}^{\rm lbe} (k)$ is negligible.
\end{lem}

{\bf Proof:} Let ${\sf negl}(k) = \left| \Pr [S_1] - \Pr [S_0] \right|$. We can easily build an adversary‎ ${\cal A}_1$ who
hopes to recover $m_b$ from Game 1. In this game, the adversary‎ ${\cal A}_1$ queries on input$(C_1 ,\,C_2 ) \ne (C_1^* ,\,C_2^* )$, while $h = h^*$. The simulator takes as input
$(C_1,C_2 ),\,h = h^*$ and computes $r = {\sf Dec}_{\rm lbe}(C_2, \cdot)\ne r^*$, $y = {C_1}/h^{*} \ne y^*$ and so $m^{'}\ne m'^{*}$. If $| m^{'}|$ is not equal to
obvious value $h^*\cdot \left\lceil {n/h^*} \right\rceil$, then the simulator rejects the ciphertext in (\ref{eq1}). Since $m'\ne m'^{*}$, thus $s = (\left|m'\right|- h^{*}\cdot v)/(k-h^*)\ne s^{*}$ and the simulator rejects the ciphertext if $s$ is not an integers. Furthermore, since the {\it position} of the message/{\sf ROB} blocks ($r \ne r^*$) and
the {\sf ROB} blocks {\it length} $s$ are not explicit, so, the output of algorithm ({\ref {alg2})) is not identical to $m_b$. Therefore, if the lattice-based encryption scheme is
secure (i.e., the adversary cannot recover $r^*$ from it), then the ${\cal A}_1$'s advantage of this game is exactly equal to ${\sf negl}(k)$. By definition of
${\sf Adv}_{{\rm \Pi}, \,{\cal A}_1}^{\rm lbe} (k)$, we have $ {\sf negl}(k) \le {\sf Adv} _{{\rm \Pi}, \,{\cal A}_1}^{\rm lbe} (k)$.\\

\begin{re} Notice that if one of the message extractor algorithm (\ref{alg2}) inputs (i.e., $r^{*}, v^{*}, s^{*}$ and $m'^*$) is not a
legitimate input, then the output of its is not identical to $m_b$.
\end{re}
\begin{re} Notice that in order to query from the simulator, the CCA2 adversary cannot modified $C_2$ based on the challenge ciphertext $C_2^*$ (well-formed decryption queries). Since for correctly retrieve $m_b$, the simulator must know the exact value of randomness $r^*$. So, if the lattice-based encryption scheme is secure, then the advantage of the CCA2 adversary is negligible.
\end{re}
{\bf Game 2.} Define Game 2 as identical with Game 1, except that $ C_1 = C_1^*$.

\begin{lem} There exists an efficient adversary ${\cal A}_2$ such that:
\begin{equation}\label{eq4}
\left|\Pr [S_2]-\Pr [S_1] \right| \le {\rm Adv} _{{\rm \Pi}, \,{\cal A}_2}^{\rm lbe} (k)
\end{equation}

By the assumption that the lattice-based encryption scheme is secure, we have that ${\sf Adv}_{{\cal A}_2}^{\rm lbe} (k)$ is negligible.
\end{lem}
{\bf Proof:} Let ${\sf negl}(k) = \left| \Pr [S_2] - \Pr [S_0] \right|$. Consider the adversary ${\cal A}_2$ who aims to recover $m_b$ from this game. In this game, the adversary
${\cal A}_2$ uniformly chooses $C_2 \ne C_2^{*}$ at random and queries on input $C=(C_1^{*},C_2)$, $h = h^{*}$. In this case, the decryption simulator computes
$r = {\sf Dec}_{\rm lbe}(C_2, \cdot)\ne r^*$. It also computes $y = {C_1}/h= y^*$, $v=v^*$, $s = s^{*}$. Although the message/{\sf ROB} blocks {\it length} and the encoded message
$m'$ are explicit, but since the {\it position} of the message/{\sf ROB} blocks are not explicit, $r \ne r^*$, thus the outputs of algorithm ({\ref {alg2})) is not identical to
$m_b$. So, if the lattice-based encryption scheme is secure, then the ${\cal A}_2$'s advantage of this game is equal to ${\sf negl}(k)$. By definition of
${\sf Adv}_{{\rm \Pi}, \,{\cal A}_2}^{\rm lbe} (k)$, we have $ {\sf negl}(k) \le {\sf Adv} _{{\rm \Pi}, \,{\cal A}_2}^{\rm lbe} (k)$.

{\bf Game 3.} Define Game 3 as identical with Game 0, except that $C_2 = C_2^*$.

\begin{lem} There exists an efficient adversary ${\cal A}_3$ such that
\begin{equation}\label{eq5}
\left| {\Pr [S_3]-\Pr [S_0]} \right| \le {\sf Adv}_{{\rm \Pi}, \,{\cal A}_3} (k).
\end{equation}
\end{lem}

{\bf Proof:} Suppose ${\sf negl}(k) = \left| \Pr [S_3] - \Pr [S_0] \right|$. We can easily build an adversary‎ ${\cal A}_3$ who
wishes to recover $m_b$ from Game 3. In this game, the adversary ${\cal A}_3$ uniformly chooses $C_1 \ne C_1^*$ at random and queries on input $(C_1 , C_2^{*})$. In this case, the
simulator computes $r = {\sf Dec}_{\rm lbe}(C_2^{*}, \cdot)= r^{*}$, $h=h^*$, $y=C_1/h^{*} \ne y^*$ and so $m'\ne m'^{*}$. If ${\sf Len}(y)=| m'|$ is not equal to obvious value
$h^*\cdot \left\lceil {n/h^*} \right\rceil$, then the simulator rejects the ciphertext in (\ref{eq1}). Since $m'\ne m'^{*}$, thus $s = (\left|m'\right|- h^{*}\cdot v)/(k-h^*)\ne s^{*}$,
and the simulator rejects the ciphertext if $s$ is not an integers. Furthermore, since the {\sf ROB} blocks {\it length} $s$ and the encoded message $m'$ are not explicit, thus the outputs
of algorithm ({\ref {alg2})) is not identical to $m_b$ and so, the ${\cal A}_3$'s advantage of this game is negligible. By definition of ${\sf Adv}_{{\rm \Pi},
\,{\cal A}_3} (k)$, we have $ {\sf negl}(k) \le {\sf Adv} _{{\rm \Pi}, \,{\cal A}_3} (k)$.\\

\begin{lem} We claim that
\begin{equation}\label{eq6}
\left| \Pr[S_3] \right| = 1/2.
\end{equation}
\end{lem}
{\bf Proof:} Game 3 same as Game 0, except that the component $C_1$ of the queried ciphertext $C=(C_1, C_2^*)$ is not computed by equation ({\ref{eq2}) but rather chosen uniformly at random. So, the queried ciphertext $C$ is statistically independent from the challenge bit $b$. Thus, the ${\cal A}_3$'s advantage in Game 3 is obviously 0, and
\[
\left| \Pr [S_3]\right| =\frac{1}{2}
\]
\textit{Completing the Proof:}

We can write
\begin{center}
$\mid \Pr [S_0] \mid=| \Pr [S_0]+\Pr [S_0]-\Pr [S_0]+\Pr [S_1]-\Pr [S_1]+\Pr [S_2]-\Pr [S_2]+$
\end{center}
\quad \quad \quad \quad \quad \quad \, $\Pr [S_3]-\Pr [S_3]|$\\
So we have
\begin{center}
 $\left| \Pr [S_0] \right| \le \left| \Pr [S_3] \right|+\left| \Pr [S_3 ] - \Pr [S_0 ] \right|+\left| \Pr [S_2 ] -\Pr [S_1 ] \right|+\left|\Pr [S_1 ] - \Pr [S_0 ] \right| +$ \\
\end{center}
\quad \quad\quad\quad \quad \,$\left| \Pr [S_2] - \Pr [S_0] \right| $\\
We have
\begin{equation} \label{eq7}
\left| \Pr [S_2 ] -\Pr [S_0 ] \right| \le \left| \Pr [S_2 ] -\Pr [S_1 ] \right|+\left| \Pr [S_1 ] -\Pr [S_0 ] \right|
\end{equation}
From equations (\ref{eq3},\ref{eq4},\ref{eq5},\ref{eq6},\ref{eq7}) we have:
\[
\left| \Pr[S_0]-1/2 \right| \le {\sf Adv}_{{\rm \Pi}, \,{\cal A}_3} (k)+2{\rm Adv} _{{\rm \Pi}, \,{\cal A}_2}^{\sf lbe} (k)+2{\rm Adv}_{{\rm \Pi}, \,{\cal A}_1}^{\sf lbe} (k)
\]
By assumption, the right-hand side of the above equation is negligible, which finishes the proof.

\subsection{Performance analysis}
The performance-related issues can be discussed with respect to the computational complexity of key generation, key sizes, encryption and decryption speed, and information
rate. The proposed cryptosystem features fast encryption and decryption. The time for computing encoded message is negligible compared to the time for computing
$({\sf Enc}_{\rm lbe}, {\sf Dec}_{\rm lbe})$. Encryption roughly needs one application of ${\sf Enc}_{\rm lbe}$ together a multiplication, and decryption roughly needs one application of ${\sf Dec}_{\rm lbe}$ together a division. The public/secret keys are as in the original scheme. The length of the ciphertext is equal to $n+(k-h)s+k$. The information rate (i.e., the ratio of the binary length of plaintext to that of the ciphertext) is equal to $n/(n+(k-h)s+k)$, and for $n\gg k$ and small integer $s$, it is close to one. Compared to other CCA2-secure lattice-based schemes were introduced today, our scheme is very simple and more efficient.

\section{Conclusion}
We construct the first direct CCA2-secure variant of the lattice-based PKE scheme, in a {\it black-box} manner, with a provable security from worst-case hardness problems in (ideal) lattices. This novel approach is very simple and more efficient and leads to the elimination of using {\sf SU-OT-SS}s or {\sf MAC}s for transformations CPA-secure schemes to a CCA2-secure one. We showed that this scheme has extra advantages, namely, its IND-CCA security remains tightly related (in the standard model) to the worst-case hardness problems in lattice. Additionally, this scheme can be used for encryption of long messages without employing the hybrid encryption method and symmetric encryption.
\begin{center}
{\bf Acknowledgment}
\end{center}
Suggestions and comments are welcome.

\end{document}